\documentstyle[11pt]{article}
\begin{document}
\baselineskip 18pt
\begin{titlepage}
\centerline{\large\bf New Constraint on the Parameters in
            Cabibbo-Kobayashi-Maskawa }
\centerline{\large\bf Matrix of Wolfenstein's Parametrization }
\vspace{1cm}

\centerline{ Yong Liu, Jing-Ling Chen }
\vspace{0.5cm}
{\small
\centerline{\bf Theoretical Physics Division}
\centerline{\bf Nankai Institute of Mathematics}
\centerline{\bf Nankai University, Tianjin 300071, P.R.China}
}
\vspace{2cm}

\centerline{\bf Abstract}

Based on the relation between CP-violation phase and the other
three mixing angles in Cabibbo-Kobayashi-Maskawa matrix postulated by us
before, a new constraint on the parameters of Wolfenstein's
parametrization is given. The result is consistent with the relative
experimental results and can be further put to the more precise
tests in future.\\\\
PACS number(s): 11.30.Er, 12.10.Ck, 13.25.+m

\vspace{2cm}
\end{titlepage}

\centerline{\large\bf New Constraint on the Parameters in
            Cabibbo-Kobayashi-Maskawa }
\centerline{\large\bf Matrix of Wolfenstein's Parametrization }
\vspace{0.5cm}

From the discovery of CP-violation by an experiment on $K^0$ decay in
1964 [1], more than thirty years have passed. Due to the physical
importance of the violation of this discrete symmetry, such as the fact
that it is one of the necessary ingredients to explain the dominance of
matter over anti-matter in our universe [2] and it could well be a
sensitive probe for new physics beyond the standard model, it has aroused
both experimental and theoretical physicists a great insterests, many
experiments have been done or are being done, number
thousands of papers about it have been published.

To understand how CP symmetry is violated, a lot of theories have been
established. The theories can be classified into three types [3,4]:
superweak [5], milliweak [6,7] which including the standard
Kobayashi-Maskawa (KM) model and millistrong [1,8]. In the
three-generation standard model, CP violation originates from the single
phase naturally occurring in the Cabibbo-Kobayashi-Maskawa (CKM)
quark-mixing matrix [9,10]. Actually, this phase is introduced somewhat
artificially, and people have been thinking it is independent of
the other three angles for many years and never doubted whether they have
a intrinsic relation [11].

Furthermore, many researchers have set up a wide variety of quark mass
matrice [12-15], the purpose is to construct the CKM matrix and to find
the relationships between quark masses and mixing angles together with the
CP violation phase. In fact, all the works can be taken as an
effort to reduce the number of free parameters presented in the standard
model, because, it is evident that the less free parameters, the less
uncertainty.

In Ref.[16], we have found that the CP-violation angle and the other
three mixing angles satisfy the following relation
\begin{equation}
sin\frac{\delta}{2}=\sqrt{\frac{sin^2\theta_1+sin^2\theta_2+sin^2\theta_3
-2(1-cos\theta_1 cos\theta_2 cos\theta_3)}{2 (1+cos\theta_1) (
1+cos\theta_2) (1+cos\theta_3)}}
\end{equation}
where $\theta_i\; (i=1,\; 2,\; 3)$ are the corresponding angles in the
standard KM parametrization matrix
\begin{equation}
V_{KM}= \left (
\begin{array}{ccc}
   c_1 & -s_1c_3& -s_1s_3 \\
   s_1c_2 & c_1c_2c_3-s_2s_3e^{i\delta}& c_1c_2s_3+s_2c_3e^{i\delta}\\
   s_1s_2 & c_1s_2c_3+c_2s_3e^{i\delta}& c_1s_2s_3-c_2c_3e^{i\delta}
\end{array}
\right )
\end{equation}
with the standard notations $s_i=sin\theta_i$ and $c_i=cos\theta_i$
are used.

The geometry meaning of Eq.(1) is very clear, $\delta$ is the solid
angle enclosed by three angles $\theta_1$, $\theta_2$ and $\theta_3$
which relate to the mixing angles among three generations,
or the area
to which the solid angle corresponds on a unit sphere. It should be noted
that, to make $\theta_1$, $\theta_2$ and $\theta_3$ enclosed a solid
angle, the condition
$$
\theta_i+\theta_j > \theta_k  \;\;\;\; (i\neq j \neq k \neq i.
\;\; i,j,k=1,2,3)
$$
must be satisfied. 

Now, we find that the CP-violation seems to originate in a geometry
reason. The deeper dynamic mechanism resulting in this kind of
geometry is not clear yet, however, we guess that it maybe
closely relate to the noncommutative rotation property of the $SU(2)$
gauge group, which describes the part of weak interaction in the
standard model. It should be refered here,
people have recognized that the CP violation parameter
$\epsilon$ is related to a certain area [15,17] more than ten years ago,
but the relation between
this area and the geometry constructed by three mixing angles
have not been recognized yet.

There are other parametrization forms of KM matrix, among them the form
introduced by Wolfenstein [18] is the most frequently used one [19-21]
in the usual references. It reads
\begin{equation}
V_{W}= \left (
\begin{array}{ccc}
   1-\frac{1}{2}\lambda^2 & \lambda & A\lambda^3(\rho-
   i\eta+i\eta\frac{1}{2}\lambda^2) \\
   -\lambda & 1-\frac{1}{2}\lambda^2-i\eta A^2 \lambda^4 &
   A\lambda^2(1+i\eta\lambda^2)\\
   A\lambda^3(1-\rho-i\eta) & -A\lambda^2 & 1
\end{array}
\right ).
\end{equation}
So, it is necessarily to convert the new constraint Eq.(1) which
represented by four angles $\delta,\; \theta_1,\; \theta_2,\; \theta_3$
into the one represented by Wolfenstein's parameters $A,\; \lambda,
 \; \rho,\;\eta$. It is easy to do so if we take use of the following
translation prescription between KM's and Wolfenstein's parameters [11]
\begin{equation}
   s_1\approx \lambda, \;\;\;\;  c_1\approx 1-\frac{\lambda^2}{2}
\end{equation}
\begin{equation}
   s_2\approx \lambda^2 A[(\rho-1)^2+\eta^2]
\end{equation}
\begin{equation}
   s_3\approx(\rho^2+\eta^2)^{1/2}A\lambda^2
\end{equation}
\begin{equation}
   sin\delta\approx \frac{\eta}{(\rho^2+\eta^2)^{1/2}}\frac{1}{
   [(\rho-1)^2+\eta^2]^{1/2}}.
\end{equation}

From Eq.(1), we obtain
\begin{equation}
sin\delta=\frac{(1+cos\theta_1+cos\theta_2+cos\theta_3)\sqrt{
sin^2\theta_1+sin^2\theta_2+sin^2\theta_3-2(1-
cos\theta_1 cos\theta_2 cos\theta_3)}}{
(1+cos\theta_1)(1+cos\theta_2)(1+cos\theta_3)}.
\end{equation}
Substituting Eqs.(4-6) to Eq.(8) and expanding the right hand side of Eq.(8)
in powers of $\lambda$, with a little more complicated calculation,
when approximate to the order of $\lambda^5$, we get
\begin{equation}
sin\delta=\frac{A \sqrt{(1-\rho)^2+\eta^2+(\rho^2+\eta^2)}}
{2 \sqrt{2}}\lambda^3
+\frac{A [(1-\rho)^2+\eta^2+(\rho^2+\eta^2)-2A^2(1-2 \rho)^2 ]}
{2^4\sqrt{2} \sqrt{(1-\rho)^2+\eta^2+(\rho^2+\eta^2)}}\lambda^5
\end{equation}
Identify the right hand sides of Eq.(9) and Eq.(7), we have
\begin{equation}
\frac{\eta}{(\rho^2+\eta^2)^{1/2}[(1-\rho)^2+\eta^2]^{1/2}}
\approx \frac{A \sqrt{(1-\rho)^2+\eta^2+(\rho^2+\eta^2)}}
{2 \sqrt{2}}\lambda^3.
\end{equation}
Here, in comparison with $\lambda^3$, we have neglected the
term of order $\lambda^5$.

Eq.(10) is the new constraint on CP-violation and quark-mixing
represented by Wolfenstein's parameters approximate to the order
of $\lambda^3$. It is the central result of this paper.

In following, we want to give a simple numerical analysis. Let
\begin{equation}
x=(\rho^2+\eta^2)^{1/2}
\end{equation}
\begin{equation}
y=[(1-\rho)^2+\eta^2]^{1/2}
\end{equation}
then
\begin{equation}
\eta=\frac{1}{2} \sqrt{2 (x^2+y^2)-(x^2-y^2)^2-1}
\end{equation}
Substituting Eqs.(11-13) to Eq.(10), we arrive
\begin{equation}
\frac{\sqrt{2 (x^2+y^2)-(x^2-y^2)^2-1}}{x y}=\frac{A \lambda^3}{\sqrt{2}}
\sqrt{x^2+y^2}
\end{equation}

Fixing $\lambda=0.22$ and $A=\frac{4}{5}$,
if we take $y=0.54\sim 1.40$ as input, then
$0.22 \sim 0.46$ for $x$ is permitted.
Hence, we find that the results are well in agreement with the experimental
analysis [22]
\begin{equation}
x=\sqrt{\rho^2+\eta^2}=0.34\pm 0.12
\end{equation}
and
\begin{equation}
y=\sqrt{(1-\rho)^2+\eta^2}=0.97\pm 0.43.
\end{equation}

In summary, we have worked out a new constraint on the parameters of
Wolfenstein's parametrization of KM matrix. Its results are consistent with
the relative experimental results and can be further put to the more precise
tests in the future.

\vspace{0.5cm}
\noindent {\bf Acknowledgment}: We would like to thank Dr. Z.Z. Xing
for helpful comments.

\end{document}